\documentclass[aps,twocolumn,showpacs,preprintnumbers,amsmath,amssymb,superscriptaddress]{revtex4}

\usepackage{graphicx}
\usepackage{dcolumn}
\usepackage{bm}

\newcommand\Tb{\mathbf{T}}
\newcommand\Pb{\mathbf{P}}
\newcommand\Qb{\mathbf{Q}}
\newcommand\pb{\mathbf{p}}
\newcommand\Bb{\mathbf{B}}
\newcommand\bb{\mathbf{b}}
\newcommand\kb{\mathbf{k}}
\newcommand\tc{\trm{c}}
\newcommand\tu{\trm{u}}
\newcommand\bs{\boldsymbol}
\newcommand\trm{\textrm}
\newcommand\apmi{\langle I\rangle}
\newcommand\napmi{{\cal I}_N}
\newcommand\iinf{{\cal I}_{\infty}}
\newcommand\nmax{n_{\trm{max}}}
\newcommand\isampl{I_{\trm{chain}}}

\begin{document}

\title{Mutual information in random Boolean models of regulatory networks}

\author{Andre S.~Ribeiro}
\affiliation{Institute for Biocomplexity and Informatics, University of Calgary, Canada}
\author{Stuart A.~Kauffman}
\affiliation{Institute for Biocomplexity and Informatics, University of Calgary, Canada}
\affiliation{Department of Physics and Astronomy, University of Calgary, Canada}
\author{Jason Lloyd-Price}
\affiliation{Institute for Biocomplexity and Informatics, University of Calgary, Canada}
\author{Bj\"orn Samuelsson}
\affiliation{Physics Department and Center for Nonlinear and Complex Systems, Duke University, Durham, NC}
\author{Joshua E.~S.~Socolar}
\affiliation{Physics Department and Center for Nonlinear and Complex Systems, Duke University, Durham, NC}

\date{\today}

\begin{abstract}
  The amount of mutual information contained in time series of two
  elements gives a measure of how well their activities are
  coordinated.  In a large, complex network of interacting elements,
  such as a genetic regulatory network within a cell, the average of
  the mutual information over all pairs $\apmi$ is a global measure of
  how well the system can coordinate its internal dynamics.  We study
  this average pairwise mutual information in random Boolean networks
  (RBNs) as a function of the distribution of Boolean rules
  implemented at each element, assuming that the links in the network
  are randomly placed.  Efficient numerical methods for calculating
  $\apmi$ show that as the number of network nodes $N$ approaches
  infinity, the quantity $N\apmi$ exhibits a discontinuity at
  parameter values corresponding to critical RBNs.  For finite systems
  it peaks near the critical value, but slightly in the disordered
  regime for typical parameter variations.  The source of high values
  of $N\apmi$ is the indirect correlations between pairs of elements
  from different long chains with a common starting point.  The
  contribution from pairs that are directly linked approaches zero for
  critical networks and peaks deep in the disordered regime.
\end{abstract}

\pacs{87.10.+e, 89.75.Fb, 02.50.Ng}


\keywords{Boolean network, Mutual Information, Criticality, Gene
Regulatory Networks}

\maketitle

\section{Introduction}\label{sec:introduction}

The dynamical behavior of a large, complex network of interacting
elements is generally quite difficult to understand in detail.  One
often has only partial information about the interactions involved and
the presence of multiple influences on each element can give rise to
exceedingly complicated dynamics even in fully deterministic systems.
A paradigmatic case is the network of genes within a cell, where the
interactions correspond to transcriptional (and post-transcriptional)
regulatory mechanisms.  The expression of a single gene may be subject
to regulation by itself and up to 15--20 proteins derived from other
genes, and the network of such interactions has a complicated
structure, including positive and negative feedback loops and
nontrivial combinatorial logic.  In this paper we study mutual
information (defined below) a measure of the overall level of
coordination achieved in models of complex regulatory networks.  We
find a surprising discontinuity in this measure for infinite systems
as parameters are varied.  We also provide heuristic explanations of
the infinite system results, the influence of noise, and finite size
effects.

The theory of the dynamics of such complicated networks begins with
the study of the simplest model systems rich enough to exhibit complex
behaviors: Random Boolean Networks (RBNs).  In a RBN model, each gene
(or ``node'') $g$ is represented as a Boolean logic gate that receives
inputs from some number $k_g$ other genes.  The RBN model takes the
network to be drawn randomly from an ensemble of networks in which
(i) the inputs to each gene are chosen at random from among all
of the genes in the system; and (ii) the Boolean rule at $g$ is
selected at random from a specified distribution over all
possible Boolean rules with $k_g$ inputs.  These two assumptions of
randomness permit analytical insights into the typical behavior of a
large network.

One important feature of RBNs is that their dynamics can be classified
as ordered, disordered, or critical.  In ``ordered'' RBNs, the
fraction of genes that remain dynamical after a transient period
vanishes like $1/N$ as the system size $N$ goes to infinity; almost
all of the nodes become ``frozen'' on an output value (0 or 1) that
does not depend on the initial state of the network
\cite{Samuelsson:ExhaustivePercol}.  In this regime the system is
strongly stable against transient perturbations of individual nodes.
In ``disordered'' (or ``chaotic'') RBNs, the number of dynamical, or
``unfrozen'' nodes scales like $N$ and the system is unstable to many
transient perturbations \cite{Samuelsson:ExhaustivePercol}.

For present purposes, we consider ensembles of RBNs parametrized by
the average indegree $K$ (i.e., average number of inputs to the nodes
in the network), and the bias $p$ in the choice of Boolean
rules.  The indegree distribution is Poissonian with mean $K$ and at
each node the rule is constructed by assigning the output for each
possible set of input values to be $1$ with probability $p$, with each
set treated independently.  If $p = 0.5$, the rule distribution is
said to be unbiased.  For a given bias, the critical connectivity,
$K_\tc$, is equal to \cite{Derrida:rand}:
\begin{equation}\label{eq:kc}
K_\tc = [2p(1 - p)]^{-1}.
\end{equation}
For $K<K_\tc$ the ensemble of RBNs is in the ordered regime; for
$K>K_\tc$, the disordered regime.  For $K=K_\tc$, the ensemble exhibits
critical scaling of the number of unfrozen nodes; e.g., the number of
unfrozen nodes scales like $N^{2/3}$.  The order-disorder transition
in RBNs has been characterized by several quantities, including
fractions of unfrozen nodes, convergence or divergence in state space,
and attractor lengths \cite{Aldana:nonlinear}.

It is an attractive hypothesis that cell genetic regulatory networks
are critical or perhaps slightly in the ordered regime
\cite{Kauffman:require,Stauffer:damage}. Critical networks display an
intriguing balance between robust behavior in the presence of random
perturbations and flexible switching induced by carefully targeted
perturbations.  That is, a typical attractor of a critical RBN is
stable under the vast majority of small, transient perturbations
(flipping one gene to the ``wrong'' state and then allowing the
dynamics to proceed as usual), but there are a few special
perturbations that can lead to a transition to a different attractor.
This observation forms the conceptual basis for thinking of cell types as
attractors of critical networks, since cell types are both homeostatic in
general and capable of differentiating when specific signals
(perturbations) are delivered.

Recently, some experimental evidence has been shown to support the
idea that genetic regulatory networks in eukaryotic cells are dynamically critical. In
Ref.~\cite{Shmulevich:critical}, the microarray patterns of gene activities
of HeLa cells were analyzed, and the trajectories in a HeLa microarray
time-series data characterized using a Lempel-Ziv complexity measure
on binarized data.  The conclusion was that cells are either ordered
or critical, not disordered. In Ref.~\cite{Ramo:aval}, it was deduced
that deletion of genes in critical networks should yield a power law
distribution of the number of genes that alter their activities with
an exponent of $-1.5$ and observed data on 240 deletion mutants in
yeast showed this same exponent.  And in Ref.~\cite{Serra:Avalanche},
micro-array gene expression data following silencing of a single gene
in yeast was analyzed.  Again, the data suggests critical dynamics for
the gene regulatory network.  These results suggest that operation at
or near criticality confers some evolutionary advantage.

In this paper we consider a feature that quantifies the sense in which
critical networks are optimal choices within the class of
synchronously updated RBNs.  We study a global measure of the
propagation of information in the network, the {\em average pairwise
  mutual information}, and show that it takes its optimal value on the
ensemble of critical networks.  Thus, within the limits of the RBN
assumptions of random structure and logic, the critical networks
enable information to be transmitted most efficiently to the greatest
number of network elements.

The average pairwise mutual information is defined as follows.  Let
$s_a$ be a process that generates a $0$ with probability $p_0$ and a
$1$ with probability $p_1$.  We define the entropy of $s_a$ as
\begin{equation}
H[s_a] \equiv -p_0\log_2 p_0 - p_1\log_2 p_1.
\end{equation}
Similarly, for a process $s_{ab}$ that generates pairs $xy$ with
probabilities $p_{xy}$, where $x,y\in\{0,1\}$, we define
the joint entropy as
\begin{align}
H[s_{ab}]& \equiv  -p_{00}\log_2 p_{00} - p_{01}\log_2 p_{01} \nonumber \\
         &  - p_{10}\log_2 p_{10} - p_{11}\log_2 p_{11}.
\end{align}

For a particular RBN, we imagine running the dynamics for infinitely
long times and starting from all possible initial configurations.  The
fraction of time steps for which the value of node $i$ is $x$ gives
$p_x$ for the process $s_i$.  The value of $p_{xy}$ for the process
$s_{ij}$ is given by the fraction of time steps for which node $i$
has the value $x$ and {\em on the next time step} node $j$ has the
value $y$.  The mutual information of the pair $ij$ is
\begin{equation}\label{eq:MIdef1}
M_{ij} = H[s_i]+H[s_j]- H[s_{ij}].
\end{equation}
With this definition, $M_{ij}$ measures the extent to which
information about node $i$ at time $t$ influences node $j$
one time step later.  Note that the propagation may be indirect; a
nonzero $M_{ij}$ can result when $i$ is not an input to $j$ but both
are influenced by a common node through previous time steps.

To quantify the efficiency of information propagation through the entire
network, we define the {\it average} pairwise mutual information for an
ensemble of networks to be
\begin{equation}\label{eq:apmi}
\apmi = \biggl\langle N^{-2}\sum_{i,j} M_{ij}\biggr\rangle ,
\end{equation}
where $\langle \cdot \rangle$ indicates an average over members of the
ensemble.  It has previously been observed that $\apmi$ is maximized
near the critical regime in numerical simulations of random Boolean
networks with a small number of nodes (less than 500)
\cite{Ribeiro:MutualInfo}.

In general, one does not expect a given element to be strongly
correlated with more than a few other elements in the network, so the
number of pairs $ij$ that contribute significantly to the sum in
Eq.~\eqref{eq:apmi} is expected to be at most of order $N$.  It is
therefore convenient to work with the quantity $\napmi \equiv N\apmi$,
which may approach a nonzero constant in the large $N$ limit.  We
use the symbol $\iinf$ to denote the $N\rightarrow\infty$ limit of $\napmi$.

As an aside, we note that other authors have considered different
information measures and found optimal behavior for critical Boolean
networks.  Krawitz and Shmulevich have found that ``basin entropy,''
which characterizes the number and sizes of basins of attraction and
hence the ability of the system to respond differently to different
inputs, is maximized for critical networks \cite{Krawitz:Basin}.
Luque and Ferrera have studied the
self-overlap~\cite{Luque:self-overlap}, which differs from $\apmi$ in
that it involves comparison of each node to its own state one time
step later (not to the state of another node that might be causally
connected to it) and that the average over the network is done before
calculating the mutual information.  Bertschinger and Natschl\"ager
have introduced the ``network-mediated separation'' ($NM$-separation)
in systems where all nodes are driven by a common input signal,
finding that critical networks provide maximal $NM$-separation for
different input signals~\cite{Bertschinger:real-time-comp}.  Our
definition of $\apmi$ places the focus on the autonomous, internal
dynamics of the network and the transmission of information along
links, which allows for additional insights into the information flow.

Two simple arguments immediately show that $\iinf$ is zero both in the
ordered regime and deep in the disordered regime.  First, note that
$M_{ij} = 0$ whenever $s_i$ or $s_j$ generates only 0s or only 1s.  In
the ordered regime, where almost all nodes remain frozen on the same
value on all attractors, the number of nonzero elements $M_{ij}$
remains bounded for large $N$.  Thus $\apmi$ must be of order $N^{-2}$
and $\iinf=0$ {\em everywhere in the ordered regime}.

Second, if $s_{ij}$ is the product of two independent processes $s_i$
and $s_j$, then $M_{ij} = 0$.  This occurs for every pair of connected
nodes in the limit of strong disorder, where $K$ is very large and the
Boolean rules are drawn from uniformly weighted distributions over all
possible rules with $k_g$ inputs.  The correlation between the output
of a node and any particular one of its inputs becomes vanishingly
small because there are many combinations of the other inputs, each
producing a randomly determined output value, so the probability for
the output to be 1 is close to $p$ for either value of the given
input.  $\iinf$ therefore vanishes in the limit of large $K$.

Given that $\iinf = 0$ for all network parameters that yield ordered
ensembles, one might expect that it rises to a maximum somewhere in
the disordered regime before decaying back to zero in the strong
disorder limit.  We show below that this is {\em not} the case.
Fixing the bias parameter $p$ at $1/2$ and allowing
the average indegree $K$ to vary, we
find that $\iinf$ exhibits a jump discontinuity at the critical value
$K=2$, then decays monotonically to zero as $K$ is increased.  The
conclusion is that among ensembles of unbiased RBNs, average pairwise
mutual information is maximized for critical ensembles.

We begin by presenting analytic arguments and numerical methods for
investigating the large system limit and establishing the existence of
a discontinuity at $K=2$ and monotonic decay
for $K>2$.  We then present results from numerical experiments
obtained by averaging over $10^{4}$ instances of networks of sizes up to $N=1000$.
These data show a strong peak near the critical value $K=2$, as
expected from the analysis.  Interestingly, the peak is substantially
higher than the size of the jump discontinuity, which may indicate
that $\iinf$ for $K=2$ is an isolated point larger than
$\lim_{K\rightarrow 2^+}\iinf$.  Finally, we present numerical results
on the variation of $\napmi$ with $p$ at fixed $K$, which again shows a
peak for critical parameter values.

\section{Average pairwise mutual information in large networks}

\subsection{Mean-field calculation of $\iinf$}

Mean-field calculations are commonly used in the theory of random
Boolean networks.  The most common forms of mean-field calculations
are within the realm of the so called annealed approximation.  In
the annealed approximation, one assumes that the rules and the inputs
are randomized at each time step.  This approach is sufficient, for
example, for calculating the average number of nodes that change value
at each time step.

For understanding the propagation of information, a slightly more
elaborate mean-field model is needed.  This mean-field model is based
on the assumption that the state of a node in a large disordered
network is independent of its state at the previous time step, but
that its rule remains fixed.  In this model, each node takes the value
$1$ with a given probability $b$, which we refer to as the local bias.
In the annealed approximation all local biases are equal because the
rules and the inputs are redrawn randomly at each time step, so the
system is characterized by a single global bias.  In our extended mean-field
model, we consider a distribution of local biases.  To determine
$\iinf$, we determine the distribution of $b$, then use it to analyze
the simple feed forward structures that provide the nonvanishing
contributions to $\iinf$ in the disordered regime.

\subsection{The distribution of local biases}

An important feature characterizing the propagation of information in
a network is the distribution of local biases.  The local bias at a
given node is determined by the rule at that node
and the local biases of its inputs.  Roughly speaking, when the bias
of the output value is stronger than the bias of the inputs,
information is lost in transmission through the node.  The local bias
distribution is defined as the self-consistent distribution obtained
as the limit of a convergent iterative process.

Let $B_t$ be the stochastic function that at each evaluation returns
a sample $b$ from the local bias
distribution at time $t$.  Then, a sample $b'$ from $B_{t+1}$ can be
obtained as follows.  Let $r$ be a Boolean rule drawn from the
network's rule distribution $R$ and let $k$ denote the number of
inputs to $r$.  Furthermore, let $\{b_1,\ldots,b_k\}$ be a set of $k$
independent samples from $B_t$.  The sample $b'$ is then given by
\begin{align}\label{eq:outbias}
  b' &= \sum_{\bs\sigma\in\{0,1\}^k} r(\bs
  \sigma)\prod_{i=1}^k[\sigma_ib_i + (1-\sigma_i)(1-b_i)]~.
\end{align}
Repeated sampling of the rule $r$ and the values $b_i$ produces 
samples $b'$ that define the distribution $B_{t+1}$.  The sequence
of distributions $B_0, B_1, B_2, \ldots$ is initiated by $B_0$ that
always returns $1/2$.

For many rule distributions $R$, $P(B_t \le x)$ converges as
$t\rightarrow\infty$.  For such rule distributions, we define $B^*$ as
the stochastic function that satisfies
\begin{align}
  P(B^* \le x) &= \lim_{t\rightarrow\infty} P(B_t \le x)
\end{align}
for all $x$.
Intuitively, $B^*$ is the large $t$ limit of $B_t$ and we use
cumulative probabilities in the definition of $B^*$ for technical
reasons: the probability density function of $B_t$ for any $t$ is a
sum of delta functions and the probability density of $B^*$ is likely
to have singularities.

We defer the evaluation of $B^*$ to Section~\ref{sec:sampling} because
some adjustments are required to obtain efficient numerical
procedures.

\subsection{\label{sec:MIfeedfw}Mutual information in feed-forward structures}

Given a rule distribution $R$ that has a well-defined distribution
$B^*$ of local biases, we calculate the mutual information between
pairs of nodes in feed-forward structures that are relevant in the
large network limit in the disordered regime.  This technique is based
on the assumption that the value of a given node at time $t + n$ is
statistically independent of the value at time $t$ for $n \ne 0$, in
which case the behavior of the inputs to a feed-forward structure can
be fully understood in terms of $B^*$.

\begin{figure}[tb]
\begin{center}
\includegraphics*[width=6.0cm]{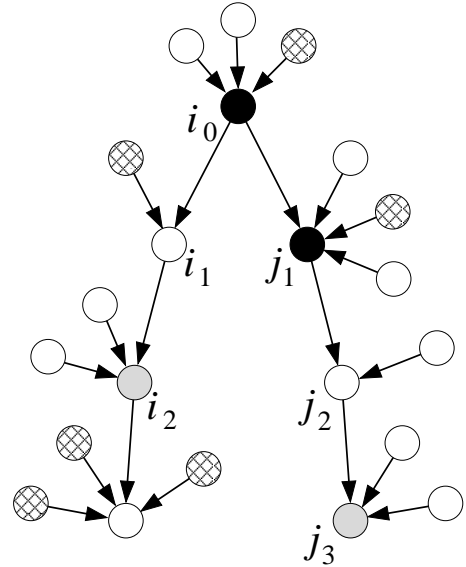}
\end{center}
\caption{Schematic structure assumed for the mean-field
  calculation of $\iinf$.  The average indegree of a node in the
  network is $K=3$.  Black nodes are an example of a directly linked
  pair.  Light grey nodes are an example of a pair that contributes to
  $\iinf$ because of a shared influence ($i_0$).  Information from
  $i_0$ takes exactly one time step longer (one additional link) to
  get to the light grey node on the right than to the one on the left.
  The node labels mark two chains of the type referred to in the text.
  Hatching indicates frozen nodes.}
\label{fig:Chains}
\end{figure}

The most direct contribution to $\langle I\rangle$ between $t$ and
\mbox{$t+1$} comes from comparing an input to a node with the output of the
same node.  Other contributions to $\langle I\rangle$ come from chains
of nodes that share a common starting point.  (See
Fig.~\ref{fig:Chains}.)  In the general case, we consider
configurations where a node $i_0$ has outputs to a chain of $n$ nodes
$i_1,\ldots,i_n$ and another chain of $n+1$ nodes
$j_1,\ldots,j_{n+1}$.  This means that node $i_m$ has one input from
node $i_{m-1}$ for $m=1,\ldots,n$, that node $j_1$ has one input from
node $i_0$, and that node $j_m$ has one input from node $j_{m-1}$ for
$m=2,\ldots,n+1$.  We allow the special case $n=0$ and let it
represent the case where an input to a node is compared to the output
of the same node.

To calculate the contribution $M_{i_n j_{n+1}}$ to $\langle I\rangle$,
we need to determine the probability distribution $p_{xy}$ for
$x=\sigma_{i_n}(t)$ and $y=\sigma_{j_{n+1}}(t+1)$.  Here we assume
that all external inputs to the feed-forward structure are
statistically independent because the probability to find a
reconnection between two paths of limited length approaches zero as
$N\rightarrow\infty$.  In Fig.~\ref{fig:Chains}, this means that there
are no (undirected) paths linking any of the pictured nodes other than
those formed by the pictured links.

Based on each rule in the structure and the biases at the external
inputs, we calculate the conditional probabilities to obtain given
output values for each value of the internal input within the
structure.  We represent this information by matrices of the form
\begin{align} \label{eq:pba}
  \Pb(\beta \mid \alpha) &\equiv
  \begin{pmatrix}
     P(\beta=0 \mid \alpha=0) & P(\beta=0 \mid \alpha=1) \\
     P(\beta=1 \mid \alpha=0) & P(\beta=1 \mid \alpha=1)
  \end{pmatrix}\,,
\end{align}
where $\alpha$ and $\beta$ are Boolean variables.  Let
\begin{align}
  \Tb_m &= \Pb\bs(\sigma_{i_m}(t)\mid \sigma_{i_{m-1}}(t-1)\bs)
  & \trm{for }m &= 1,\ldots,n\,,\\
  \Tb'_1 &= \Pb\bs(\sigma_{j_1}(t)\mid \sigma_{i_0}(t-1)\bs)\,, & \trm{and}\\
  \Tb'_m &= \Pb\bs(\sigma_{j_m}(t)\mid \sigma_{j_{m-1}}(t-1)\bs) &
  \trm{for }m &= 2,\ldots,n+1\,.
\end{align}
Note that the elements in each of these matrices depend on the rule
chosen at the node index in the first argument of $\Pb$ and that the choice
of rule specifies the number $k$ of inputs to that node.
Multiplication of these matrices corresponds to following a signal
that passes through the feed-forward structure, so we have
\begin{align}
  \Pb\bs(\sigma_{i_n}(t + n)\mid \sigma_{i_0}(t)\bs) &= \Tb_n\Tb_{n-1}\cdots\Tb_1
\intertext{and}
  \Pb\bs(\sigma_{j_{n+1}}(t + n + 1)\mid \sigma_{i_0}(t)\bs) &=
                                      \Tb'_{n+1}\Tb'_n\cdots\Tb'_1\,.
\end{align}

The probabilities for the pairs $\sigma_{i_n}(t) \sigma_{j_{n+1}}(t+1)$
can be expressed as elements of a matrix
$\Pb\bs(\sigma_{i_n}(t),\sigma_{j_{n+1}}(t+1)\bs)$ where
\begin{align}
  \Pb(x, y) &\equiv
  \begin{pmatrix}
     P(x = 0, y = 0) & P(x = 0, y = 1) \\
     P(x = 1, y = 0) & P(x = 1, y = 1)
  \end{pmatrix}\,.
\end{align}
Note that $\Pb(x,y)$ is the matrix of values $p_{xy}$ defined
previously, which has a different meaning from $\Pb(\beta\mid\alpha)$.
In accordance with the definition of $M_{ij}$ above, we define the mutual
information associated with a matrix $\Qb$ with elements $q_{xy}$ to be
\begin{align}\label{eq:IofP}
I(\Qb) &= \sum_{x,y} q_{xy} \log_2\frac{q_{xy}}{\bigl(\sum_z q_{xz}\bigr)
\bigl(\sum_z q_{zy}\bigr)}\,.
\end{align}

Now let $\Pb_n$ denote $\Pb\bs(s_{i_n}(t), s_{j_{n+1}}(t+1)\bs)$ and
let
\begin{align}
  \Bb_0 &\equiv
  \begin{pmatrix}
     1 - b_{i_0} & 0 \\
     0 & b_{i_0}
  \end{pmatrix}\,.
\end{align}
We can then write
\begin{align}\label{eq:Pn}
  \Pb_n &= \Tb_n\Tb_{n-1}\cdots\Tb_1\Bb_0(\Tb'_1)^\top(\Tb'_2)^\top
              \cdots(\Tb'_{n+1})^\top\,.
\end{align}

For a given set of the indegrees $k_{i_1},\ldots,k_{i_n}$ and
$k_{j_1},\ldots,k_{j_{n+1}}$ denoted by $\kb$, we let $\langle
I(\Pb_n)\rangle_{\kb}$ denote the average mutual information
associated with $\Pb_n$.  Note that $\langle I(\Pb_n)\rangle_{\kb}$ is
the contribution to $\iinf$ arising from the average over $i_n
j_{n+1}$ pair of nodes in chains with a given $\kb$.

The average number of occurrences of a feed-forward
structure with the vector $\kb$ is given by $Nw_\kb$, where
\begin{align} \label{eq:wk}
  w_\kb &= \prod_{m=1}^{2n+1} k_mP(k = k_m)
\end{align}
and $P(k = k_m)$ denotes the probability that a randomly selected rule
from the rule distribution $R$ has $k_m$ inputs.

Putting together Eqs.~\eqref{eq:pba}--\eqref{eq:wk}, we obtain the
expression
\begin{align}\label{eq:limNI}
  \iinf &=
      \sum_{n=0}^\infty\sum_{\kb\in\mathbb Z_+^{2n+1}}w_\kb
      \langle I(\Pb_n)\rangle_{\kb}
\end{align}
for rule distributions in the disordered regime.

Numerical evaluation of this expression is cumbersome, but can be
streamlined substantially by handling the frozen nodes (local biases
$b=0$ and $b=1$) analytically.  Removing these nodes from the core of
the calculations provides additional insights and yields a version of
Eq.~\eqref{eq:limNI} that can be sampled more efficiently by
Monte-Carlo techniques.

The fraction of unfrozen nodes is given by $u = P(B^*\notin {0,1})$
and the distribution of local bias in the set of unfrozen nodes is
given by $B^*_\tu$, where one sample $b$ of $B^*_\tu$ is obtained by
sampling $b$ from $B^*$ repeatedly until a value of $b$ not equal to 0
or 1 is obtained.  Similarly, we define an altered rule distribution
$R_\tu$.  A sample $r_\tu$ from $R_{\rm u}$ is obtained by sampling $r$ from
$R$ and fixing each input to $0$ with probability $P(B^*=0)$ and $1$
with probability $P(B^*=1)$.  New samples $r$ are drawn until one
obtains a nonconstant function $r_\tu$ of the $k_\tu$ inputs that are
not frozen.  (The probability of an unfrozen node having a given value
of $k_\tu$ is given below.)

We define $\Pb_n^\tu$ and $w_{\kb}^\tu$ by replacing $B^*$ and $R$ with
$B^*_\tu$ and $R_\tu$ in the definitions of $\Pb_n$ and $w_{\kb}$,
respectively.  With these definitions, we rewrite Eq.~\eqref{eq:limNI}
and get
\begin{align}\label{eq:limNIu}
  \iinf &=
      u\sum_{n=0}^\infty\sum_{\kb\in\mathbb Z_+^{2n+1}}w_\kb^\tu
      \langle I(\Pb_n^\tu)\rangle_{\kb}\,.
\end{align}

Let $R(K)$ denote the rule distribution for a Poissonian distribution
of indegrees $k$ such that $\langle k\rangle=K$ and uniform
distributions among all Boolean rules with a given $k$.  Note that the
definition does not require that a rule does depend on all of its
inputs.  From this point, we restrict the discussion to distributions
of the form $R(K)$.  We expect the qualitative behavior of this rule
distribution to be representative for a broad range of rule
distributions.  The critical point for $R(K)$ occurs at $K = 2$, with
ordered networks arising for $K < 2$ and disordered networks for $K >
2$.

The symmetric treatment of 0s and 1s in the rule distribution
simplifies the calculation of $u$ in the sense that it is sufficient
to keep track of the probability to obtain constant nodes from $B_t$
and there is no need to distinguish nodes that are constantly 1 from
those that are constantly 0.  Let $u_t$ denote $P(B_t \notin\{0,1\})$.
Then, $u_{t+1}$ can be calculated from $u_t$ according to
\begin{align}\label{eq:umap}
   u_{t+1} &= e^{-Ku_t}\sum_{k=0}\frac{(Ku_t)^k}{k!}\bigl(1 - 2^{2^k-1}\bigr)\,.
\end{align}
The desired value $u$ is given by the stable fixed point of the map
$u_t \mapsto u_{t+1}$.   Note that this map is identical to the damage
control function presented in Ref.~\cite{Samuelsson:ExhaustivePercol},
meaning that the above determination of $u$ is consistent with the
process of recursively identifying frozen nodes without the use of any
mean-field assumption.

The rule distribution $R_\tu (K)$ within the set of unfrozen nodes gives
a rule with $k_m$ inputs with probability
\begin{align}\label{eq:kudistr}
  P(k_\tu = k_m) &=  e^{-Ku}\frac{(Ku)^{k_m}}{u\,k_m !}\bigl(1 - 2^{2^{k_m}-1}\bigr)
\end{align}
The distribution of rules with $k$ inputs is uniform among all
nonconstant Boolean rules with the given number of inputs.  This
expression can be used in Eq.~\eqref{eq:wk}.  Further analysis is
helpful in determining the limiting value of $\iinf$ as $K$ approaches
its critical value of 2 from above.  Appendix \ref{app:crit} addresses
this issue.

\subsection{\label{sec:sampling}Numerical sampling}

We are now in a position to evaluate Eq.~\eqref{eq:limNIu} by an
efficient Monte-Carlo technique.  To obtain a distribution that
is a good approximation of $B_\tu^*$, we use an iterative process to
create vectors of samples.  Each vector has a fixed number $S$ of
samples.  The process is initiated by a vector $\bb_0$ where all $S$
nodes are set to $1/2$.  Then a sequence of vectors is created by
iteratively choosing a vector $\bb_{t+1}$ based on the previous vector
$\bb_t$.  To obtain each node in $\bb_{t+1}$, we use
Eq.~\eqref{eq:outbias} with $r$ sampled from $R'_\tu(K)$ defined below
and $b_1,\ldots,b_k$ being randomly selected elements of $\bb_t$.

If the rule distribution $R'_\tu(K)$ were set to $R_\tu(K)$, the
sequence of vectors would fail to converge properly for $K$ that are
just slightly larger than $2$.  For such $K$, $R_\tu(K)$ gives a
1-input rule with a probability close to 1.  This leads to slow
convergence and to a proliferation of copies of identical bias values
in the sequence of vectors $\{\bb_t\}$.  The remedy for this problem
is quite simple.  Due to the symmetry between 0 and 1 in the rule
distribution, application of a 1-input rule to an input bias
distribution $B^*$ gives the same output bias distribution $B^*$.
Thus, we can remove the 1-input rules from $R_\tu(K)$ without altering
the limiting distribution at large $t$.  We let $R'_\tu(K)$ denote the
rule distribution obtained by disregarding all 1-input samples
from $R_\tu(K)$.

Based on $\{\bb_t\}$, we construct matrices that can be used
for estimating the sums in Eqs.~\eqref{eq:limNIu} and
\eqref{eq:limKNI} by random sampling.  After an initial number of
steps required for convergence, we sample $S$ matrices of the form
$\Pb\bs(r(\bs\sigma)\mid \sigma_1\bs)$ where $r$ is drawn from $R'_\tu(K)$ and
the inputs $\sigma_2,\ldots,\sigma_k$ have biases drawn from $\bb_t$.
These matrices and the indegrees of the corresponding rules are
stored in the vectors $\pb_t$ and $\kb_t$, respectively.  The elements
of the vectors $\bb_t$, $\pb_t$, and $\kb_t$ are indexed by $i =
1,\ldots,S$ and the $i$th element of each vector is denoted by
$b_{t,i}$, $\pb_{t,i}$, and $k_{t,i}$, respectively.  For notational
convenience, we define
\begin{align}
  \pb_{t,0} &= \begin{pmatrix}1&0\\0&1\end{pmatrix}
\quad{\rm and}\quad
  k_{t,0} = 1
\end{align}
for all $t = 0,1,2,\ldots$.  With these definitions, the $i=0$
elements correspond to a copy operator.

To estimate the sum in Eq.~\eqref{eq:limNIu}, we truncate it at $n =
\nmax$ where $\nmax$ is chosen to be sufficiently large for the
remaining terms to be negligible.  Then we obtain random samples in
the following way.  We select $i_0$ uniformly from $\{1,\ldots,S\}$
and set each of the indices $i_1,\ldots,i_{\nmax},
j_1,\ldots,i_{\nmax+1}$ to $0$ with probability $P(k_\tu = 1)$ or to a
uniformly chosen sample of $\{1,\ldots,S\}$ with probability $P(k_\tu
> 1)$.  Then we set
\begin{align}
 & \Pb^\tu_0 = \begin{pmatrix} 1 - b_{i_0} & 0 \\ 0 & b_{i_0} \end{pmatrix}
     (\pb_{t,j_1})^\top\,,\quad
  \kappa_0 = k_{t,j_1}\,;
\intertext{and}
 & \left.\begin{array}{l}
  \Pb^\tu_n = \pb_{t,i_n}\Pb^\tu_{n-1}(\pb_{t,j_{n+1}})^\top \\
  \kappa_n = k_{t,i_n}\kappa_{n-1}k_{t,j_{n+1}}
  \end{array}\right\}
  \trm{for }n = 1,\ldots,\nmax.
\end{align}
With $I(\Pb^\tu_n)$ given by Eq.~\eqref{eq:IofP}, we construct
samples of the mutual information associated with sets of nodes
in chain structures:
\begin{align}
  \isampl &= \sum_{n=0}^{\nmax} \kappa_n I(\Pb^\tu_n)\,.
\end{align}
The average value of $\isampl$ provides an approximation of the sum
in Eq.~\eqref{eq:limNIu} and we get
\begin{align}
  \iinf &\approx u\langle\isampl\rangle\,.
\end{align}
This approximation is good if $t$, $\nmax$, and $S$ are sufficiently
large.  For evaluating $\langle\isampl\rangle$, we draw $S$ samples of
$\isampl$ for several subsequent $t$ that are large enough to ensure
convergence in $\bb_t$.

The technique just described is easily generalized to account for
uncorrelated noise in the dynamics.  To model a system in which each
node has a probability $\epsilon$ of generating the wrong output at
each time step, we need only modify $r(\bs\sigma)$ in Eq.~\eqref{eq:outbias} and
$\Pb(\beta\mid\alpha)$ of Eq.~\eqref{eq:pba} as follows:
\begin{align} \label{eq:outbiasnoise}
   r_{\epsilon}(\bs\sigma) & = (1-\epsilon)\,r(\bs\sigma)
          + \epsilon\,[1-r(\bs\sigma)]
 \intertext{and}
 \label{eq:pbanoise}
  \Pb_{\epsilon}(\beta \mid \alpha) & =
  \begin{pmatrix}
     1-\epsilon & \epsilon \\
     \epsilon & 1-\epsilon
  \end{pmatrix}
  \Pb(\beta \mid \alpha)\,,
\end{align}
where $\Pb$ is determined as above by the Boolean rule at a given
node.  For nonzero $\epsilon$, all nodes are unfrozen, but the
calculations proceed exactly as above with $b'$ and $\Pb$ replaced by
$b'_{\epsilon}$ and $\Pb_{\epsilon}$, respectively.

We use a similar technique to estimate $\lim_{K\rightarrow2_+}$ based
on Eq.~\eqref{eq:limKNI}.  The main differences in this technique are
that 1-input nodes do not enter the numerical sampling and the sum to
be evaluated is two-dimensional rather than one-dimensional.

\section{\label{sec:results}Results}

\subsection{The large system limit}

\begin{figure}[tb]
\begin{center}
\includegraphics*[width=\columnwidth]{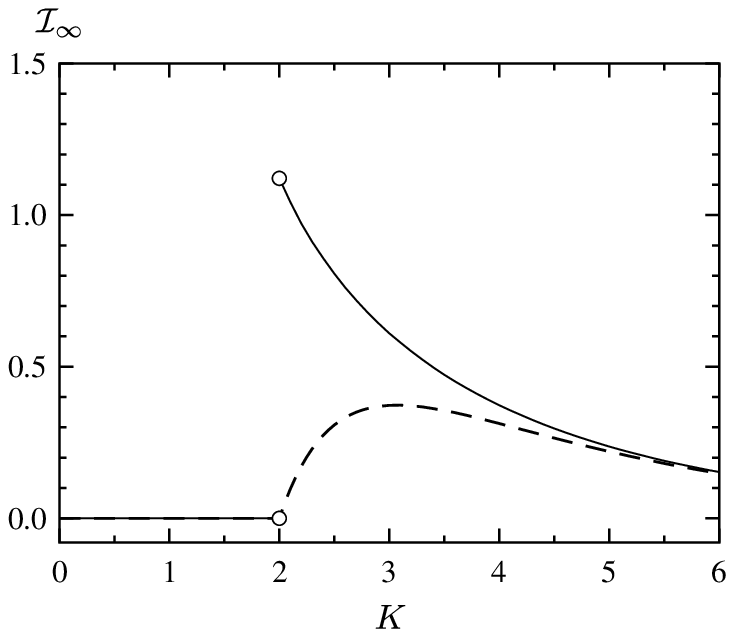}
\end{center}
\caption{The large system limit $\iinf$ for $N\apmi$ (solid line) and
  the contribution to $\iinf$ from direct information transfer through
  single nodes (dashed line).  The empty circles at the discontinuity
  of $\iinf$ indicate that we do not know the value of $\iinf$ for
  $K=2$.  The size of the sample vectors is $S=10^4$.  The number of
  vectors used was $10^3$--$10^4$ and these were drawn after $10^3$
  steps taken for convergence in $\bb_t$. The summation cutoff $\nmax$
  varies from $6$ for high $K$ to $100$ for $K$ close to 2.  For the
  limit of $\iinf$ for $K\rightarrow 2_+$ using Eq.~\eqref{eq:limKNI},
  increasing the summation cutoff from $\nmax=20$ to $\nmax=40$ gave
  no significant difference in the result.}
\label{fig:I_inf}
\end{figure}

Using the the expressions derived in Section~\ref{sec:MIfeedfw} and
the stochastic evaluation techniques described in
Section~\ref{sec:sampling}, we have obtained estimates of $\iinf$ for
$K>2$.  Using the expressions in Appendix~\ref{app:crit}, we obtain
$\lim_{K\rightarrow 2_+}\iinf$.  The results are shown in
Fig.~\ref{fig:I_inf}.

The solid line in Fig.~\ref{fig:I_inf} shows the full result for
$\iinf$.  The dashed line shows the contribution to $\iinf$ that comes
from pairs of nodes that are directly linked in the network.  It is
interesting to note that the direct links alone are not responsible
for the peak at criticality.  Rather, it is the correlations between
indirectly linked nodes that produce the effect, and in fact
dominate $\iinf$ for $K$ at and slightly above the critical value.

\begin{figure}[tb]
\begin{center}
\includegraphics*[width=\columnwidth]{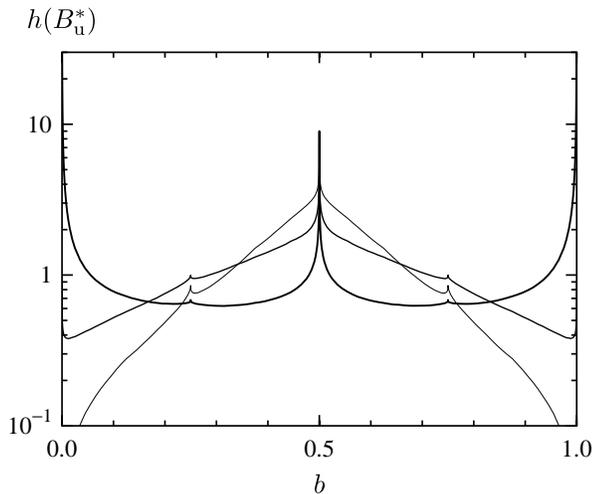}
\end{center}
\caption{Histograms $h(B^*_\tu)$ of the distributions of unfrozen
  local biases $b$
  drawn from $B^*_\tu$ for $K\rightarrow2_+$ (bold line), $K=3$
  (medium line), and $K=4$ (thin line).  Bins of width $10^{-4}$ were
  used to estimate the probability density from a sequence of $10^6$
  sample vectors $\bb_t$ that were drawn after $10^3$ steps for
  convergence.  The size of the sample vectors is $S=10^4$.  The
  combination of a small bin-width and a large sample size enables a
  clear picture of the strongest singularities.}
\label{fig:bias}
\end{figure}

The distribution of local biases plays an important role in
determining $\iinf$.  Biases that are significantly different from $b
= 1/2$ are important for $K$ that are not deep into the disordered
regime, and the distribution of local biases is highly nonuniform.
Dense histograms of biases drawn from the distribution $B^*_\tu$ for
various $K$ are shown in Fig.~\ref{fig:bias}.  Singularities at $b=0$
and $b=1$ occur for $K$ in the range $2 < K \lesssim 3.4$, and for all
$K > 2$ there is a singularity at $b = 1/2$.

\begin{figure}[tb]
\begin{center}
\includegraphics*[width=\columnwidth]{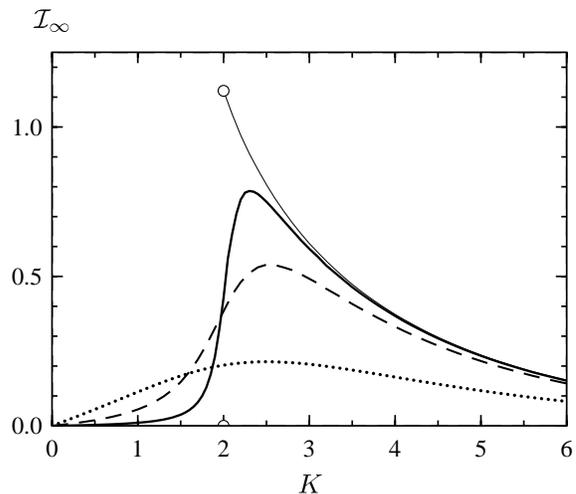}
\end{center}
\caption{The large system limit $\iinf$ as a function of $K$ for
  various noise levels $\epsilon$ in the updating.  The thin solid
  line shows $\iinf$ for networks without noise as displayed in
  Fig.~\ref{fig:I_inf}.  The other lines represent $\epsilon = 0.001$
  (thick solid line), $0.01$ (dashed line), and $0.1$ (dotted line).
  The size of the sample vectors is $S=10^4$.  $10^3$--$10^6$
  were drawn after $10^3$ steps taken for convergence in $\bb_t$.
  Extensive sampling was required close to criticality for
  $\epsilon = 0.001$.}
\label{fig:I_inf_noise}
\end{figure}

When uncorrelated noise is added to each node at each time step,
$\iinf$ may decrease due to the random errors, but may also {\em
  increase} due to the unfreezing of nodes.  The net effect as a
function of $K$ is shown in Fig.~\ref{fig:I_inf_noise} for the case
where each output is inverted with probability $\epsilon$ on each time
step.  As $\epsilon$ is increased from zero, the peak shifts to the
disordered regime and broadens.  The mutual information due to random
unfreezing is clearly visible on the ordered side.  In the regime
where indirect contributions dominate $\iinf$, however, there is a
strong decrease as correlations can no longer be maintained over long
chains.  Deep in the disordered regime, we see the slight decrease
expected due to the added randomness.  For $\epsilon \gtrsim 0.1$, the
maximum of $\iinf$ shifts back toward $K=2$.  In fact, it can be shown
that as $\epsilon$ approaches $1/2$, which corresponds to completely
random updating, the $\iinf$ curve approaches
\begin{equation}\label{eq:Iinf_noisy}
\iinf = \frac{K}{\ln 2}\,\biggl(\frac{1}{2}-\epsilon\biggr)^{\!2}\exp(-K/2)\,.
\end{equation}
In this limit, the maximum occurs at $K=2$ and the peak height scales
like $(1/2-\epsilon)^2$.  The fact that the critical $K$ is recovered
in the strong noise limit is coincidental; it would not occur for
most other choices of Boolean rule distributions.

\subsection{Finite size effects}

Numerical simulations on finite networks reveal an important feature
near the critical value of $K$ that is not analytically accessible
using the above techniques because of the difficulty of calculating
$\iinf$ right at the critical point.  (We have only computed the limit
as $K$ approaches $K_\tc$, not the actual value at $K_\tc$.)  We compute
$\apmi$ by sampling the mutual information from pairs of nodes from
many networks.

In collecting numerical results to compare to the $\iinf$ calculation,
there are some subtleties to consider.  The calculations are based on
correlations that persist at long times in the mean-field model.  To
observe these, one must disregard transient dynamics and also average
over the dynamics of different attractors of each network.  The latter
average should be done by including data from all the attractors in
the calculation of the mutual information, {\em not} by calculating
separate mutual information calculated for individual attractors.  For
the results presented here, we have observed satisfactory convergence
both for increasing lengths of discarded transients and for increasing
numbers of initial conditions per network.  Finally, an accurate
measurement of the mutual information requires sufficiently long
observation times; short observation times lead to systematic
overestimates of the mutual information. (See, for example,
Ref.~\cite{Bazso:MI}.) In the figures below, the size of the spurious
contribution due to finite observation times is smaller than the
symbols on the graph.

\begin{figure}[tb]
\begin{center}
\includegraphics*[width=\columnwidth]{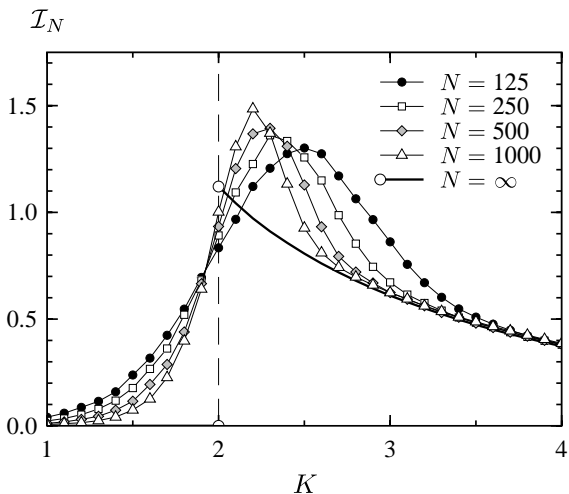}
\end{center}
\caption{\protect $\napmi$ as a function of $K$ for several different
  system sizes.  For these calculation we use $10^4$ networks with
  $40$ runs from different initial states per network and a discarded
  transient of length $10^4$ updates for each run.  (For large $K$,
  good convergence was obtained for discarded transients of length
  $10^3$.)  The sequences of states were recorded for a sample of
  $10N$ pairs of nodes in each network. The vertical dashed line
  indicates the critical value of $K$.} \label{fig:finiteK}
\end{figure}

Fig.~\ref{fig:finiteK} shows that the peak in $\napmi$ extends well
above the computed $\iinf$ value.  The figure shows $\napmi$ as a
function of $K$ for several system sizes $N$.  As $N$ increases, the
curve converges toward the infinite $N$ value both in the ordered and
disordered regimes.  In the vicinity of the critical point, however,
the situation is more complicated.  The limiting value at criticality
will likely depend on the order in which the large size and
$K\rightarrow K_\tc$ limits are taken.

\begin{figure}[tb]
\begin{center}
\includegraphics*[width=\columnwidth]{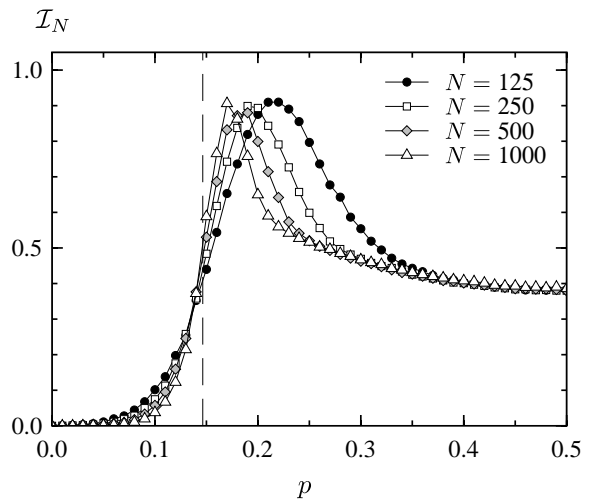}
\end{center}
\caption{\protect $\napmi$ as a function of $p$ for several different
  system sizes.  For these calculation we use $10^4$ networks with
  $40$ runs from different initial states per network and a discarded
  transient of length $10^4$.  (For some data points far into the
  disordered regime, good convergence was obtained for discarded
  transients of length $10^3$.)  The sequences of states were recorded
  for a sample of $10N$ pairs of nodes in each network.  The vertical
  dashed line indicates the critical value of $p$.}
\label{fig:finitep}
\end{figure}

We have also studied $\napmi$ as a function of the bias parameter $p$,
while holding $K$ fixed at $4$.  Fig.~\ref{fig:finitep} shows that
$\napmi$ is again peaked at the critical point $p=(2-\sqrt{2})/4$; the
qualitative structure of the curves is the same as that for varying
$K$.  The calculation of $\iinf$ for $p\neq 1/2$ requires modifications
of the analysis described above that are beyond the scope of this work.

\section{Special rule distributions}

Up to now, the discussion has focused on rule distributions
parametrized only by an independent probability $p$ of finding a $1$
in a given row of the truth table for any given node.  Consideration
of other possibilities shows that $\iinf$ can actually be made as
large as desired in networks that are as deep as desired in the
disordered regime.  Let $\lambda$ be the average sensitivity of a node
to its inputs; i.e., the average number of nodes that change values
when the value of one randomly selected node is flipped.  $\lambda=1$ is
one criterion for identifying critical networks
\cite{Shmulevich:Activities}.  For any value of
$\lambda$ in the disordered regime ($\lambda>1$) and any target value
${\cal I}$ of $\iinf$, one can always define a rule distribution that
gives a random network characterized by $\lambda$ and ${\cal I}$.  The
key to constructing the distribution is the observation that long
chains of single-input nodes produce large $\iinf$ and that a small
fraction of nodes with many inputs and maximally sensitive rules
(multi-input versions of {\sc xor}) is enough to make $\lambda$ large.

Consider the following class of random networks.  Each node has an
indegree $k$ of either $1$ or $g$, with the probability of having $g$
inputs being $\gamma$.  The logic function at each node is the parity
function or its negation.  For $k=1$ nodes this means they either
copy or invert their input. (There are no nodes with constant
outputs.)  For $k=g$ nodes it means that a change in any single input
causes a change in the output.  Note that there are no frozen nodes in
these networks.

For these networks, we have
\begin{align}
\lambda = \langle k\rangle = 1-\gamma + g\gamma.
\end{align}
The network consists of $\gamma N$ nodes with multiple inputs, which
can be thought of as the roots of a tree of single-input nodes.  If
$g^2 \ll \gamma N$, loops in the graph will be rare enough that they
will have little effect on the average pairwise mutual information.
If $g$ and $\gamma$ are fixed and $N$ is taken to infinity, loops can
be neglected in computing $\iinf$.  For a node with $g>1$, the mutual
information between any given input node and the output is zero for
the rule distribution under consideration.  This is because the bias
distribution in networks consisting entirely of maximally sensitive
nodes is a delta function at $b=1/2$.  Thus $\langle
I(\Pb_n)\rangle_{\kb} = 0$ for all $\kb \neq \{1,1,\ldots ,1\}$.  For
$\kb = \{1,1,\ldots ,1\}$, Eq.~\eqref{eq:wk} gives $w_\kb =
(1-\gamma)^{2n+1}$ and we get from Eq.~\eqref{eq:limNI}:
\begin{align}
  \iinf = \frac{1 - \gamma}{\gamma (2-\gamma)}.
\end{align}
By choosing $\gamma \ll 1$ and $g \gg 1/\gamma$ we can make $\iinf$ as
large as desired while simultaneously making $\lambda$ as large as
desired.

Generalization of this construction to networks with a broader
distribution of indegrees and/or rules is straightforward.  Roughly
speaking, high $\iinf$ occurs deep in the disordered regime when there
is a small fraction of
nodes of high indegree and high sensitivity and the remaining nodes
are sensitive to exactly one input.

\section{Conclusions}

In the introduction above, we noted early evidence that eukaryotic
cells may be dynamically critical.  Our calculations indicate that,
{\em within the class of RBNs with randomly assigned inputs to each node and
  typically studied rule distributions}, critical networks provide an
optimal capacity for coordinating dynamical behaviors.  This type of
coordination requires the presence of substantial numbers of dynamical
(unfrozen) nodes, the linking of those nodes in a manner that allows
long-range propagation of information while limiting interference from
multiple propagating signals, and a low error rate.  To the extent
that evolutionary fitness depends on such coordination and RBN models
capture essential features of the organization of genetic regulatory
networks, critical networks are naturally favored.  We conjecture that
mutual information is optimized in critical networks for broader
classes of networks that include power-law degree distributions and/or
additional local structure such as clustering or over-representation of
certain small motifs.

A key insight from our study is that the maximization of average
pairwise mutual information is achieved in RBNs by allowing long
chains of effectively single-input nodes to emerge from the background
of frozen nodes and nodes with multiple unfrozen inputs.  The
correlations induced by these chains are reduced substantially when
stochastic effects are included in the update rules, thus destroying
the jump discontinuity in $\iinf$ at the critical point and shifting
the curve toward the dashed one in Fig.~\ref{fig:I_inf} obtained from
direct linkages only.  Though the noise we have modeled here is rather
strong, corresponding to a large fluctuation in the expression of a
given gene from its nominally determined value, a 
shift of the maximum into the disordered regime may be expected to occur in other models.

The behavior of the average pairwise mutual information in RBNs with
flat rule distributions is nontrivial and somewhat surprising.  This
is due largely to the fact that the network of unfrozen nodes in
nearly critical systems does indeed have long single-input chains.  By
choosing a rule distribution carefully, however, we can arrange to
enhance the effect and produce arbitrarily high values of $\iinf$ even
deep in the disordered regime.  Whether real biological systems have
this option is less clear.  The interactions between transcription
factors and placement of binding sites required to produce logic with
high sensitivity to many inputs appear difficult (though not impossible) to
realize with real molecules \cite{Hwa:logic}.

Maximization of pairwise mutual information may be a sensible proxy
for maximization of fitness within an ensemble of evolutionarily
accessible networks: we suggest that systems based on high-$\apmi$
networks can orchestrate complex, timed behaviors, possibly allowing
robust performance of a wide spectrum of tasks.  If so, the
maximization of pairwise mutual information within the space of
networks accessible via genome evolution may play an important role in natural
selection of real genetic networks.  We have found that maximization
of pairwise mutual information can be achieved deep in the disordered
regime by sufficiently nonuniform Boolean rule distributions.
However, in the absence of further knowledge, a roughly flat rule
distribution remains the simplest choice, and in this case pairwise
mutual information is maximized for critical networks.
Given the tentative evidence for criticality in real genetic
regulatory networks
\cite{Shmulevich:critical,Ramo:aval,Serra:Avalanche}, these
results may be biologically important.


\begin{acknowledgments}
  We thank M.~Andrecut for stimulating
  conversations.  This work was supported by the National Science
  Foundation through Grant No.~PHY-0417372 and by the Alberta
  Informatics Circle of Research Excellence through Grant No.~CPEC29.
\end{acknowledgments}

\appendix

\section{\label{app:crit}Approaching criticality}

The mean-field calculations in Section~\ref{sec:MIfeedfw} are not
applicable to critical networks, but we can investigate the behavior
for disordered networks that are close to criticality.  In this
appendix, we investigate the limit
\begin{align}
   \lim_{K\rightarrow2_+}\iinf
\end{align}
for networks with the rule distribution $R(K)$.

Let $K=2+\epsilon$ where $\epsilon$ is a small positive number.  The
fraction of unfrozen nodes goes to zero as $\epsilon \rightarrow 0$,
meaning that it is appropriate to expand Eq.~\eqref{eq:umap} for small
$u_t$.  A second order Taylor expansion yields
\begin{align}
  u_{t+1} &\approx \tfrac12Ku_t - \tfrac1{16}K^2u_t^2
\end{align}
and the fixed point $u$ satisfies
\begin{align}
  1 &\approx \tfrac12K - \tfrac1{16}K^2u\,,
\end{align}
meaning that
\begin{align}
  u &\approx 8\frac{K-2}{K^2} \approx 2\epsilon\,.
\end{align}
Approximation of Eq.~\eqref{eq:kudistr} to the same order gives
\begin{align}
  P(k_\tu = 1) &\approx 1 - \tfrac72\epsilon\\
\intertext{and}\label{eq:Pu2}
  P(k_\tu = 2) &\approx \tfrac72\epsilon\,.
\end{align}
The probability to obtain $k_\tu>2$ vanishes to the first order in
$\epsilon$.  Equation~\eqref{eq:Pu2} yields that $P(k_\tu = 2) \approx
\tfrac74u$ for small $\epsilon$.  However, all rules with $k_\tu = 2$
are not proper 2-input rules in the sense that they do not depend on both
inputs.  Of the 14 nonconstant Boolean 2-input
rules, 4 are dependent on only one input and are effectively 1-input
rules.  Hence, the probability $p_2$ for an unfrozen node to have a proper
2-input rule is given by
\begin{align}\label{eq:p2}
  p_2 &\approx \tfrac54u\,.
\end{align}

For small $\epsilon$, nodes with single inputs dominate the expression
for $\Pb_n$ in \eqref{eq:Pn}.  A matrix $\Tb$ corresponding to a
one-input rule is either the unity matrix or a permutation matrix that
converts 0s to 1s and vice versa in the probability
distribution.  None of these matrices has any effect on $\langle
I(\Pb_n)\rangle$ or $\langle I(\Pb_n^\tu)\rangle$, because of the
symmetry between 0s and 1s in the rule distribution.  Hence, we can
express $\langle I(\Pb^\tu_n)\rangle_\kb$ on the form
\begin{align}
   \langle I(\Pb^\tu_n)\rangle_\kb &= \langle
                      I(\Pb^\tu_{(n'_0,n'_1)})\rangle_{\kb'}\,,
\end{align}
where $n'_0$ and $n'_1$, respectively, are the numbers of indegrees
$k_{i_1}, \ldots, k_{i_n}$ and $k_{j_1}, \ldots, k_{j_{n+1}}$ that are
different from $1$ and $\kb'$ is corresponding vector of indegrees
different from $1$.

In the limit $\epsilon\rightarrow0_+$, we can neglect indegrees larger
than $2$.  Hence, we introduce $\langle I(\Pb_{n_0,n_1}^{(2)})\rangle$
to denote the average mutual information of
\begin{align}\label{eq:P2nn}
  \Pb^{(2)}_{(n_0,n_1)}
       &= \Tb_n\Tb_{n-1}\cdots\Tb_1\Bb_0(\Tb'_1)^\top(\Tb'_2)^\top
              \cdots(\Tb'_{n_1})^\top\,
\end{align}
where $\Tb_1,\ldots,\Tb_n$ and $\Tb'_1,\ldots,\Tb'_{n'}$ correspond to
randomly selected 2-input rules that do depend on both inputs.  Both
$\Bb_0$ and the $\Tb$-matrices are drawn based on the distribution of
local biases obtained by proper 2-input rules only, because the symmetry
between 0s and 1s ensures that rules with one input do not alter the
equilibrium distribution $B_\tu^*$.

Then, Eq.~\eqref{eq:limNIu} can be approximated by
\begin{align}
  \iinf &\approx
      u \sum_{n=0}^\infty\sum_{n_0=0}^n\sum_{n_1=0}^{n+1}
         \binom n{n_0}\binom{n + 1}{n_1}
      \langle I(\Pb_{n_0,n_1}^{(2)})\rangle
        \nonumber\\
      &\phantom{\approx} \times
        (2p_2)^{n_0 + n_1}(1 - p_2)^{2n + 1 - n_0 - n_1}
\end{align}
for small $\epsilon$.  This approximation is exact in the limit
$\epsilon\rightarrow0_+$ and reordering of the summation gives
\begin{align}
  \lim_{\epsilon\rightarrow 0_+}\iinf
    &= \sum_{n_0,n_1\in\mathbb{N}^2} W_{(n_0,n_1)}
      \langle I(\Pb_{n_0,n_1}^{(2)})\rangle\,,
\end{align}
where
\begin{align}
  W_{(n_0,n_1)} &\equiv \lim_{\epsilon\rightarrow0_+}
   u\sum_{n=0}^\infty\binom n{n_0}\binom{n + 1}{n_1}
        \nonumber\\
      &\phantom{\approx} \times
        (2p_2)^{n_0 + n_1}(1 - p_2)^{2n + 1 - n_0 - n_1}\\
   &=  \lim_{\epsilon\rightarrow0_+} \frac{u}{2p_2}\binom{n_0+n_1}{n_0}\,.
\end{align}
Because $\lim_{\epsilon\rightarrow0_+}u/(2p_2) = 2/5$, we get
\begin{align}\label{eq:limKNI}
   \lim_{K\rightarrow2_+}\iinf
    &= \frac25\sum_{n_0,n_1\in\mathbb{N}^2} \binom{n_0+n_1}{n_0}
      \langle I(\Pb_{n_0,n_1}^{(2)})\rangle\,.
\end{align}

For approaching the critical point from the ordered regime, we know
from the discussion in Section~\ref{sec:introduction} that
\begin{align}
   \lim_{K\rightarrow2_-}\iinf &= 0\,,
\end{align}
meaning that $\iinf$ has a discontinuity at $K=2$.  From the scaling
of the number of unfrozen nodes and the number of relevant nodes, we
expect that $\iinf$ is well-defined and different from 0 for $K=2$ but
we have found no analytical hints about whether this value is larger
or smaller than $\lim_{K\rightarrow2_+}\iinf$.

Numerical evaluation of the sum in Eq.~\eqref{eq:limKNI} is
carried out in close analogy with the technique described in
Section~\ref{sec:sampling}.

\bibliography{asrBibMI}

\begin{thebibliography}{15}
\expandafter\ifx\csname natexlab\endcsname\relax\def\natexlab#1{#1}\fi
\expandafter\ifx\csname bibnamefont\endcsname\relax
  \def\bibnamefont#1{#1}\fi
\expandafter\ifx\csname bibfnamefont\endcsname\relax
  \def\bibfnamefont#1{#1}\fi
\expandafter\ifx\csname citenamefont\endcsname\relax
  \def\citenamefont#1{#1}\fi
\expandafter\ifx\csname url\endcsname\relax
  \def\url#1{\texttt{#1}}\fi
\expandafter\ifx\csname urlprefix\endcsname\relax\def\urlprefix{URL }\fi
\providecommand{\bibinfo}[2]{#2}
\providecommand{\eprint}[2][]{\url{#2}}

\bibitem[{\citenamefont{Samuelsson and
  Socolar}(2006)}]{Samuelsson:ExhaustivePercol}
\bibinfo{author}{\bibfnamefont{B.}~\bibnamefont{Samuelsson}} \bibnamefont{and}
  \bibinfo{author}{\bibfnamefont{J.~E.~S.} \bibnamefont{Socolar}},
  \bibinfo{journal}{Phys.\ Rev.\ E} \textbf{\bibinfo{volume}{74}},
  \bibinfo{pages}{036113} (\bibinfo{year}{2006}).

\bibitem[{\citenamefont{Derrida and Pomeau}(1986)}]{Derrida:rand}
\bibinfo{author}{\bibfnamefont{B.}~\bibnamefont{Derrida}} \bibnamefont{and}
  \bibinfo{author}{\bibfnamefont{Y.}~\bibnamefont{Pomeau}},
  \bibinfo{journal}{Europhys. Lett.} \textbf{\bibinfo{volume}{1}},
  \bibinfo{pages}{45} (\bibinfo{year}{1986}).

\bibitem[{\citenamefont{Aldana-Gonzalez
  et~al.}(2003)\citenamefont{Aldana-Gonzalez, Coppersmith, and
  Kadanoff}}]{Aldana:nonlinear}
\bibinfo{author}{\bibfnamefont{M.}~\bibnamefont{Aldana-Gonzalez}},
  \bibinfo{author}{\bibfnamefont{S.}~\bibnamefont{Coppersmith}},
  \bibnamefont{and} \bibinfo{author}{\bibfnamefont{L.~P.}
  \bibnamefont{Kadanoff}}, \emph{\bibinfo{title}{{\normalfont in}
  {P}erspectives and Problems in Nonlinear Science{\normalfont, edited by E.
  Kaplan, J. E. Marsden and K. R. Sreenivasan}}} (\bibinfo{publisher}{Springer,
  New York}, \bibinfo{year}{2003}), p.~\bibinfo{pages}{23}.

\bibitem[{\citenamefont{Kauffman}(1990)}]{Kauffman:require}
\bibinfo{author}{\bibfnamefont{S.~A.} \bibnamefont{Kauffman}},
  \bibinfo{journal}{Physica D} \textbf{\bibinfo{volume}{42}},
  \bibinfo{pages}{135} (\bibinfo{year}{1990}).

\bibitem[{\citenamefont{Stauffer}(1994)}]{Stauffer:damage}
\bibinfo{author}{\bibfnamefont{D.}~\bibnamefont{Stauffer}},
  \bibinfo{journal}{J. Stat. Phys.} \textbf{\bibinfo{volume}{74}},
  \bibinfo{pages}{1293} (\bibinfo{year}{1994}).

\bibitem[{\citenamefont{Shmulevich et~al.}(2005)\citenamefont{Shmulevich,
  Kauffman, and Aldana}}]{Shmulevich:critical}
\bibinfo{author}{\bibfnamefont{I.}~\bibnamefont{Shmulevich}},
  \bibinfo{author}{\bibfnamefont{S.~A.} \bibnamefont{Kauffman}},
  \bibnamefont{and} \bibinfo{author}{\bibfnamefont{M.}~\bibnamefont{Aldana}},
  \bibinfo{journal}{Proc.\ Natl.\ Acad.\ Sci.\ USA}
  \textbf{\bibinfo{volume}{102}}, \bibinfo{pages}{13439}
  (\bibinfo{year}{2005}).

\bibitem[{\citenamefont{R{\"a}m{\"o} et~al.}(2006)\citenamefont{R{\"a}m{\"o},
  Kesseli, and Yli-Harja}}]{Ramo:aval}
\bibinfo{author}{\bibfnamefont{P.}~\bibnamefont{R{\"a}m{\"o}}},
  \bibinfo{author}{\bibfnamefont{J.}~\bibnamefont{Kesseli}}, \bibnamefont{and}
  \bibinfo{author}{\bibfnamefont{O.}~\bibnamefont{Yli-Harja}},
  \bibinfo{journal}{J.\ Theor.\ Biol.} \textbf{\bibinfo{volume}{242}},
  \bibinfo{pages}{164} (\bibinfo{year}{2006}).

\bibitem[{\citenamefont{Serra et~al.}(2004)\citenamefont{Serra, Villani, and
  Semeria}}]{Serra:Avalanche}
\bibinfo{author}{\bibfnamefont{R.}~\bibnamefont{Serra}},
  \bibinfo{author}{\bibfnamefont{M.}~\bibnamefont{Villani}}, \bibnamefont{and}
  \bibinfo{author}{\bibfnamefont{A.}~\bibnamefont{Semeria}},
  \bibinfo{journal}{J.\ Theor.\ Biol.} \textbf{\bibinfo{volume}{227}},
  \bibinfo{pages}{149} (\bibinfo{year}{2004}).

\bibitem[{\citenamefont{Ribeiro et~al.}(2006)\citenamefont{Ribeiro, Este,
  Lloyd-Price, and Kauffman}}]{Ribeiro:MutualInfo}
\bibinfo{author}{\bibfnamefont{A.~S.} \bibnamefont{Ribeiro}},
  \bibinfo{author}{\bibfnamefont{R.~A.} \bibnamefont{Este}},
  \bibinfo{author}{\bibfnamefont{J.}~\bibnamefont{Lloyd-Price}},
  \bibnamefont{and} \bibinfo{author}{\bibfnamefont{S.~A.}
  \bibnamefont{Kauffman}}, \bibinfo{journal}{WSEAS Trans.\ on Systems}
  \textbf{\bibinfo{volume}{5}}, \bibinfo{pages}{2935} (\bibinfo{year}{2006}).

\bibitem[{\citenamefont{Krawitz and Shmulevich}(2007)}]{Krawitz:Basin}
\bibinfo{author}{\bibfnamefont{P.}~\bibnamefont{Krawitz}} \bibnamefont{and}
  \bibinfo{author}{\bibfnamefont{I.}~\bibnamefont{Shmulevich}},
  \bibinfo{journal}{Phys.\ Rev.\ Lett.} \textbf{\bibinfo{volume}{98}},
  \bibinfo{pages}{158701} (\bibinfo{year}{2007}).

\bibitem[{\citenamefont{Luque and Ferrera}(2000)}]{Luque:self-overlap}
\bibinfo{author}{\bibfnamefont{B.}~\bibnamefont{Luque}} \bibnamefont{and}
  \bibinfo{author}{\bibfnamefont{A.}~\bibnamefont{Ferrera}},
  \bibinfo{journal}{Complex Systems} \textbf{\bibinfo{volume}{12}},
  \bibinfo{pages}{241} (\bibinfo{year}{2000}).

\bibitem[{\citenamefont{Bertschinger and
  Natschl\"ager}(2004)}]{Bertschinger:real-time-comp}
\bibinfo{author}{\bibfnamefont{N.}~\bibnamefont{Bertschinger}}
  \bibnamefont{and}
  \bibinfo{author}{\bibfnamefont{T.}~\bibnamefont{Natschl\"ager}},
  \bibinfo{journal}{Neural Comput.} \textbf{\bibinfo{volume}{16}},
  \bibinfo{pages}{1413} (\bibinfo{year}{2004}).

\bibitem[{\citenamefont{Bazs\'{o} et~al.}(2004)\citenamefont{Bazs\'{o},
  Zal\'{a}nyi, and Petr\'{o}czi}}]{Bazso:MI}
\bibinfo{author}{\bibfnamefont{F.}~\bibnamefont{Bazs\'{o}}},
  \bibinfo{author}{\bibfnamefont{L.}~\bibnamefont{Zal\'{a}nyi}},
  \bibnamefont{and}
  \bibinfo{author}{\bibfnamefont{A.}~\bibnamefont{Petr\'{o}czi}},
  \bibinfo{journal}{Proc.\ IEEE Intl.\ Joint Conf.\ on Neural Net.}
  \textbf{\bibinfo{volume}{4}}, \bibinfo{pages}{2843} (\bibinfo{year}{2004}).

\bibitem[{\citenamefont{Shmulevich and Kauffman}(2004)}]{Shmulevich:Activities}
\bibinfo{author}{\bibfnamefont{I.}~\bibnamefont{Shmulevich}} \bibnamefont{and}
  \bibinfo{author}{\bibfnamefont{S.~A.} \bibnamefont{Kauffman}},
  \bibinfo{journal}{Phys. Rev. Lett.} \textbf{\bibinfo{volume}{93}},
  \bibinfo{pages}{048701} (\bibinfo{year}{2004}).

\bibitem[{\citenamefont{Buchler et~al.}(2003)\citenamefont{Buchler, Gerland,
  and Hwa}}]{Hwa:logic}
\bibinfo{author}{\bibfnamefont{N.~E.} \bibnamefont{Buchler}},
  \bibinfo{author}{\bibfnamefont{U.}~\bibnamefont{Gerland}}, \bibnamefont{and}
  \bibinfo{author}{\bibfnamefont{T.}~\bibnamefont{Hwa}},
  \bibinfo{journal}{Proc.\ Natl.\ Acad.\ Sci.\ USA}
  \textbf{\bibinfo{volume}{100}}, \bibinfo{pages}{5136} (\bibinfo{year}{2003}).

\end{thebibliography}

\end{document}